\documentclass[final,3p,times,twocolumn]{elsarticle}

\usepackage{amssymb}

\usepackage{tabularx}

\def\ileak{$I_{\rm leak}$}
\journal{Nuclear Instruments and Methods in Physics Research A}

\begin{document}

\begin{frontmatter}



\title{Extensive testing of Schottky CdTe detectors for the ECLAIRs X-/Gamma-ray Camera on board the SVOM mission}


\author{Remou\'e N., Barret D., Godet O., Mandrou P.}

\address{Universit\'e de Toulouse (UPS), 118 route de Narbonne 31062 Toulouse Cedex 9, France \\
Centre National de la Recherche Scientifique, Centre d'Etude Spatiale des Rayonnements, UMR 5187, 9 av. du Colonel Roche, BP 44346, 31028 Toulouse Cedex 4, France }

\begin{abstract}
We report on an on-going test campaign of more than 5000 Schottky CdTe detectors ($4\times4\times1$ mm$^3$), over a sample of twelve thousands, provided by Acrorad Co., Ltd (Japan). 6400 of these detectors will be used to build the detection plane of the ECLAIRs camera on the Chinese-French gamma-ray burst mission SVOM. These tests aim at eliminating, from the flight model, detectors showing high leakage current and unstable behaviors. They are mandatory to fulfill the prime requirement of ECLAIRs to detect gamma-ray burst photons down to 4 keV. For better functional performance and stability, the detectors will be operated at $-20^\circ$C under a reverse bias of 600 V. Under these conditions, we found that 78\% of the detectors already tested could be considered for the flight model. They show a low initial leakage current (with a mean value around 20 pA), and remain stable below 100 pA within two hours. Using a standard analog electronics chain, we measured a mean energy resolution of 1.8 keV at 59.6 keV using an $^{241}$Am source. 
Because the Schottky detectors are well known to be unstable due to the bias-induced polarization effect, the high voltage power supplies on ECLAIRs will have to be switched off at regular time intervals. We investigated the polarization effect first at room temperature and low bias voltage for faster analysis. We found that the spectroscopic degradation in quantum efficiency, gain and energy resolution, starts as soon as the bias is turned on: first slowly and then dramatically after a time $t_p$ which depends on the temperature and the voltage value. Preliminary tests under in-flight conditions ($-20^\circ$C, -600 V) showed that the detectors should remain stable over a timescale larger than a day. 
As a by product of our test campaign, we measured the mean activation energy of 170 Schottky CdTe detectors. We found evidence for two distinct populations of detectors: the main one centered at 0.64 eV, interpreted as due to cadmium vacancies in the crystal, and the second population centered at 0.54 eV, correlated with a lower apparent resistivity.
\end{abstract}


\end{frontmatter}

\section{Introduction}
The SVOM (\textit{Space-based multi-band astronomical Variable Objects Monitor}) gamma-ray burst mission will carry the ECLAIRs 2D coded-mask telescope, operating in the 4-250 keV range \cite{2009AIPC.1133...25G,2009AIPC.1133...76R}. CdTe semi-conductor detectors are well suited for detecting X- and gamma-rays in this energy range, thanks to the high detection efficiency of CdTe, due to a high atomic number ($Z_{\rm Cd}$=48, $Z_{\rm Te}$=52) and a high density ($\rho_{\rm CdTe}$ = 5.85 g.cm$^{-3}$) \cite{2001ITNS...48..950T}. The 4 keV low-energy threshold is a driver for the mission, as it shall improve significantly the sensitivity of ECLAIRs to high redshift gamma-ray bursts \cite{2009NIMPA.603..365G}. For this purpose, the front-end electronics integrated in the ASIC IDeF-X \cite{2006NIMPA.567..140G,2009ITNS...56.2351G} for Cd(Zn)Te detectors has been chosen. It shows a low intrinsic noise (Equivalent Noise Charge (ENC) floor of 33 e$^{-}$ rms at 6 $\mu$s peaking time), low consumption ($\sim 3$ mW/channel) and is radiation hard (Single Event Latchup Linear Energy Transfer threshold of 56 MeV.cm$^{2}$.mg$^{-1}$). However, it is suited only to low noise detectors and low capacitive materials. Therefore, we must choose detectors with the lowest leakage current, in order to minimize the parallel noise.

\begin{figure}[!h]
\begin{center}
\centerline{\includegraphics[width=.45\textwidth]{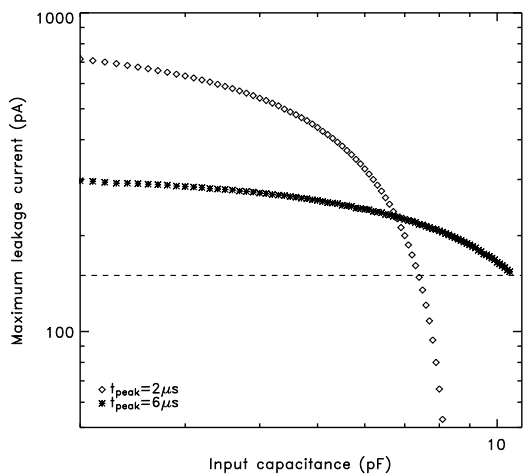}}
  \caption{The maximum leakage current accepted to ensure a 4 keV low energy threshold (defined at 6 sigma of the centre of the noise peak) as a function of the input capacitance (detector + parasitic). Two values are compared for the peaking time: 2 $\mu$s and 6 $\mu$s. We used the theoretical ENC expression and measured noise parameters of the ASIC IDeF-X ECLAIRs \cite{2006NIMPA.567..140G,2009ITNS...56.2351G}. The 150 pA limit is shown for indication (dashed line).}
\label{remoue:fig1}
\end{center}
\end{figure}

Given the theoretical expression of the ENC at the output of the shapers of the ASIC \cite{2006NIMPA.567..140G}, while using measured values of the noise coefficients $\alpha_{\rm d}$ and $\alpha_{\rm 1/f}$ from IDeF-X ECLAIRs (table II in \cite{2009ITNS...56.2351G}), we can simulate the evolution of the ENC as a function of the detector leakage current ($I_{\rm leak}$). These values depend on both the peaking time chosen (between 2 and 6 $\mu$s typically), and the total capacitance at the input of the charge sensitive amplifier of the ASIC. The later may vary between 2 pF and 10.5 pF \cite{2009ITNS...56.2351G}. It includes the detector capacitance together with the parasitic capacitance, mainly due to interconnections between the detector and the electronics (ceramic boards in our case). Assuming a 6 sigma low-energy threshold, and considering the worst case (i.e. 10.5 pF input capacitance) and 6 $\mu$s peaking time, detecting photons of 4 keV requires \ileak\ to be lower than 150 pA during the in-flight operation of the camera (Figure \ref{remoue:fig1}).

Such a low value of \ileak\ could be achieved with ohmic CdTe detectors operated at low temperature, as were previously used for the ISGRI hard X-ray imager aboard INTEGRAL \cite{2003A&A...411L.141L}, but only under the condition of a low bias voltage (because of the linear increase of \ileak\ with the bias in this case). But lower bias voltage means less charge collected, because of the low mobility and short lifetime of holes in CdTe \cite{2001ITNS...48..950T}. Therefore, Schottky CdTe diodes are favoured. Indeed, the Schottky blocking junction enables to apply higher bias voltage without increasing too much the leakage current, which in turn improves the charge collection efficiency, reducing the distortion of the X-ray spectrum. The detection plane of ECLAIRs will then be an assembly of 6400 Schottky CdTe monolithic pixels with a  $4\times4$ mm$^{2}$ area and 1 mm thickness provided by Acrorad Co., Ltd (Japan) \cite{2009AIPC.1133...76R}. The anode is made of Indium and the cathode of Platinum. The crystal is grown by the so-called Traveling Heater Method (THM), and is doped with Cl \cite{1999NIMPA.436..120F}. This is the first time such planar monolithic In/CdTe/Pt detectors are considered for hard X-ray imaging in space, as ISGRI aboard INTEGRAL used Pt/CdTe/Pt detectors \cite{2003A&A...411L.141L} and the BAT telescope on \textit{Swift} used CdZnTe pixels \cite{2005SSRv..120..143B}. More recently, pixellated devices are considered as for the Caliste 64 micro-camera \cite{2009NIMPA.610..154M} for the Symbol-X mission, or the Si/CdTe Compton camera \cite{2007NIMPA.579..871W} on board Astro-H.

Unfortunately, Schottky-type CdTe detectors suffer from the bias-induced polarization phenomenon, which manifests itself by a decrease in quantum efficiency, a degradation of the spectral resolution and a shift of the photo-peak towards lower energies \cite{1974ITNS...21..322M,2006JaJAP..45.8842T,2009ApPhL..94j2113C,2007ApPhL..90f3504O}. A way to minimize and delay considerably this effect is to operate the detectors at low temperature and high bias voltage \cite{2006JaJAP..45.8842T,1998ITNS...45..428M}. Thus ECLAIRs will keep an in-flight nominal temperature of $-20^\circ$C $\pm$ $2^\circ$C over the whole detection plane, with the detectors reverse biased at 600 V. Moreover, this effect being reversible, a way to maintain stable performance of the detectors over time is to turn off the high voltage at regular time intervals. In the current baseline configuration, the high voltage is expected to be switched off once a day for ECLAIRs, when the spacecraft will pass in the deepest part of the South Atlantic Anomaly. 
Another type of polarization induced by high photon fluence irradiation has been reported for Cd(Zn)Te detectors,  typically above $10^{5}$ photons/mm$^{2}$ integrated over about 100 ms \cite{1988NIMPA.263..457V,1999MedPh..26...38J,2008ITNS...55..572S}. It shall not be an issue for ECLAIRs, as an on-axis exceptionally bright gamma-ray burst should generate less than 1 photon/mm$^{2}$/100 ms.

Thirty two pixels are to be mounted on a monolithic Al$_2$O$_3$ ceramic plate and hybridized with the ASIC IDeF-X ECLAIRs, to form the so-called XRDPIX elementary module \cite{2009AIPC.1133...76R,remoue09}. In order to minimize the parallel noise of the ASIC, we decided to select the most suitable CdTe detectors to be used for the flight before mounting them on each XRDPIX. In fact, each channel of the ASIC has its own adjustable discrimination threshold \cite{2009ITNS...56.2351G}. But, because we aim for homogeneous modules, we considered during the selection process that the performance will be dictated by the worst detector on a given XRDPIX. In total, more than twelve thousands will be tested. Their leakage currents at two operating temperatures ($-20^\circ$C and $+25^\circ$C) will be measured and their spectroscopic performance assessed. 

In this paper, we first report on the results of a uniform testing of more than 5000 detectors, which led to the selection criteria to be applied to the flight detectors. Albeit similarly large numbers of detectors currently flowing on INTEGRAL and \textit{Swift} have been tested \cite{1999NIMPA.428..216L,2005NIMPA.541..372S}, this is the first time Schottky CdTe detectors are tested so extensively. We then describe the results of complementary tests of a few selected samples of detectors showing either acceptable performance or atypical behaviors during the automatic tests; the latter could only be identified because of the large sample of detectors considered. Those tests include the measurement of the polarization effect, as well as the time evolution of the dark current, and finally the measurement of the detector performance as a function of the operating temperature, in particular to define a precise value of their activation energy. Those tests are required to anticipate the detector behavior in space and optimize the in-flight performance of the ECLAIRs camera, under the general constraints of the SVOM mission (low-earth orbit, frequent South Atlantic Anomaly passages) and science requirements (detection of GRB photons down to 4 keV).
\label{intro}

\begin{figure}
\begin{center}
\centerline{\includegraphics[width=.49\textwidth]{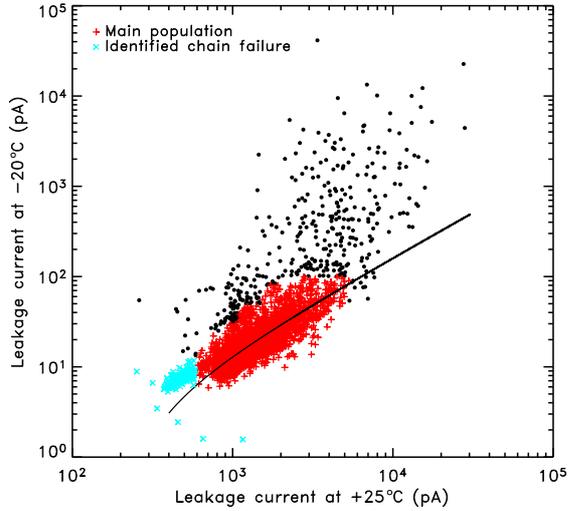}}
  \caption{The leakage current measured at $-20^\circ$C against the leakage current at $+25^\circ$C, in pA, for 5000 detectors (at -600 V). The cluster of plus signs identifies the main population of detectors, containing about 90\% of the total sample. A linear fit is shown for indication (a=$1.6\times10^{-2}$, b=$-3.5$ pA). Detectors deviating from the main correlation line are considered suspicious; some of them have been further investigated (see Figure \ref{remoue:fig4}). The cluster of cross marks at the bottom left of the plot (less than 2.5\% of the total number of detectors tested) can be explained by the aging of the electronic chain.}
\label{remoue:fig2}
\end{center}
\end{figure}

\begin{figure}[!h]
\begin{center}
 \centerline{\includegraphics[width=.49\textwidth]{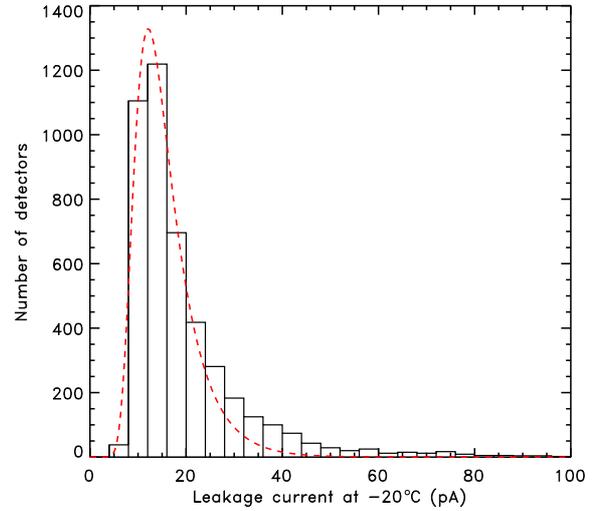}}
       \caption{The histogram of leakage currents measured at $-20^\circ$C (-600 V) for the main population (corresponding to the red cluster in Figure \ref{remoue:fig2}). For indication, the histogram was fitted by a Landau function, which peaks at 12 pA. The distribution at $+25^\circ$C shows the same pattern, with a peak value at 1.1 nA.}
\label{remoue:fig3}
\end{center}
\end{figure}

\section{Functional performance of Schottky CdTe detectors}
In order to test more than twelve thousand detectors, at $-20^\circ$C and $+25^\circ$C, a dedicated automatic test bench has been developed for both leakage current measurements and spectroscopy. The experimental setup is described in \cite{remoue09}, with the ambiant hygrometry ($\le$ 50\%) and temperature ($\pm 0.4^\circ$C) controlled all along the procedure. The $+25^\circ$C level is useful for two reasons. First, we can compare our values with Acrorad measurements; an important deviation between the two independent measurements could be indicative of detector damage or a failure in our measuring chain. Secondly, functional tests will be performed all along the instrument integration at room temperature (or slightly cooler). Therefore, we need to know what parallel noise level to expect at such a temperature. Before testing, each detector is visually inspected (peculiar marks or chipouts are listed) and then placed inside a shield box, thus remaining in the dark 24 hours before the first test is performed. Each day, 32 detectors are simultaneously tested, requiring the presence of two technicians over regular working hours. The test bench has been running continuously since October 2008 and more than 6500 detectors have been tested up to December 2009. The results presented in this section concern the first 5000 detectors tested.

\subsection{Leakage current measurements}
As for the leakage current measurements, the CdTe are polarized with a reverse bias of 600 V, and once the operating temperature is stabilized, a measurement of \ileak\ is recorded with a picoammeter (Keithley 6487) 30 seconds after the high voltage is set. Figure \ref{remoue:fig2} shows the variation of this initial \ileak\ at $-20^\circ$C against \ileak\ measured at $+25^\circ$C, together with a linear fit for the main population inferred from the plot (identified by plus signs). The histogram of the leakage current measured at $-20^\circ$C is shown on Figure \ref{remoue:fig3}. The main population, which contains about 90\% of the total sample, shows a mean initial leakage current of 19 pA at $-20^\circ$C and 1.4 nA at $+25^\circ$C, with a standard deviation of 12 pA and 0.6 nA respectively. As a first step of our selection procedure, from this plot (Figure \ref{remoue:fig2}), one can already identify suspicious detectors, for which the two leakage currents deviate significantly from the correlation line fitting the main population. 
The cluster of detectors at the bottom left of Figure \ref{remoue:fig2} (cross marks), which represents less than 2.5\% of the detectors tested, have lower \ileak\ at $+25^\circ$C. They were all measured over a continuous time period, suggesting a failure of our measurement chain. It was found that one component of the chain (a few resistances in the high voltage supplier dispatcher) broke down. After the chain was repaired, a sample of those detectors was tested again; none were found with the same low \ileak\ and all returned on the main correlation line. 

\begin{figure}
\begin{center}
 \centerline{\includegraphics[width=.49\textwidth]{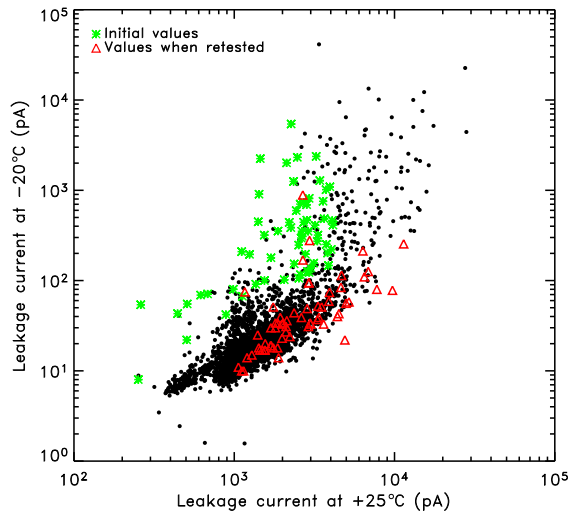}}
       \caption{Same as figure \ref{remoue:fig2}. Those detectors away from the main correlation line (marked as stars) have been tested a second time. For most of them, the new values of leakage currents (marked as empty triangles) are found near the main correlation line, thus indicating that their abnormal values during the first set of tests likely reflected a degraded experimental set-up (contamination by residual humidity). }
\label{remoue:fig4}
\end{center}
\end{figure}

The remaining population, spreading over two decades of currents at $-20^\circ$C, has also been further investigated. We have tested a second time a sample of 64 detectors located away from the main group in Figure \ref{remoue:fig2} (they are marked as stars in Figure \ref{remoue:fig4}). 90\% of them are now found in the main group (marked as empty triangles in Figure \ref{remoue:fig4}). We explained this discrepancy between the two measurements as probably due to residual humidity trapped inside the shield box containing the detector during the first test. This would naturally explain why the effect is most visible at $-20^\circ$C than at $+25^\circ$C. Furthermore, the CdTe diodes we use are not equipped with a guard ring \cite{2004ITNS...51.1881N}, thus they are more sensitive to edge and surface currents caused by excessive humidity. For the remaining 10\%, still away from the main group after the second test, a visual inspection did not reveal any peculiar surface degradation or alteration. Similarly their performance, as previously measured by Acrorad at $+25^\circ$C, were found acceptable. Therefore, intrinsic defects of the crystal might be involved, and those detectors will be discussed in section 4.

In any case, it is not currently possible to re-test all the detectors falling outside the main correlation. So by safety, all the detectors failing the acceptance level of \ileak\ for the ASIC, and located more than three sigma away from the main correlation line after the first test, are not considered for the flight model of ECLAIRs.

\subsection{Evaluation of spectroscopic performance}
\begin{figure}
\begin{center}
\centerline{\includegraphics[width=.45\textwidth]{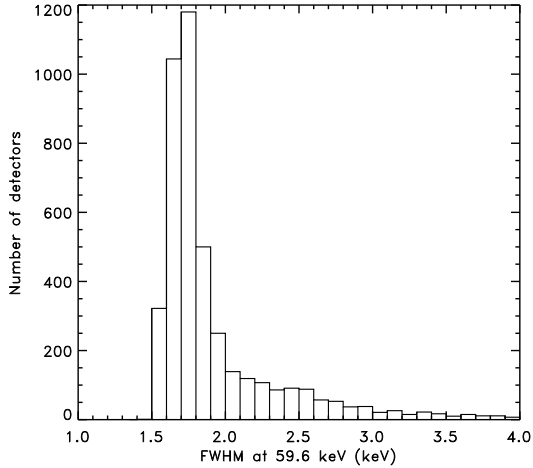}}
  \caption{Histogram of the energy resolution (FWHM) at 59.6 keV as computed from $^{241}$Am X-ray spectrum ($-20^\circ$C, 600 V reverse bias) for the main population of the 5000 detectors tested, i.e. those showing acceptable $I_{\rm leak}$. The best fit Landau function peaks at 1.7 keV. A similar spectral resolution is measured at 13.9 keV. About 10\% detectors are not in the core of the distribution.}
\label{remoue:fig5}
\end{center}
\end{figure}

\begin{figure}
\begin{center}
\centerline{\includegraphics[width=.45\textwidth]{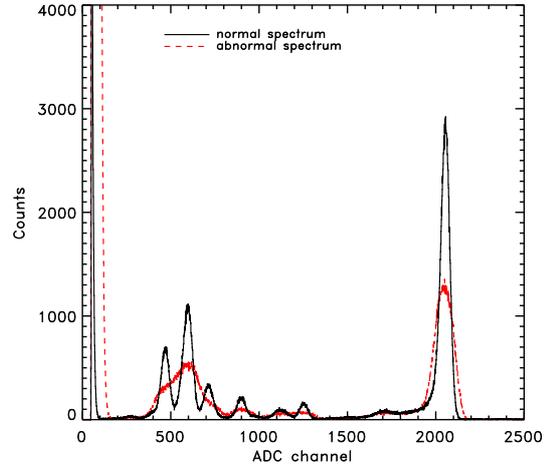}}
  \caption{The $^{241}$Am spectrum for a good (solid line) and a bad (dashed line) detector respectively, measured at $-20^\circ$C (-600 V). Both were found acceptable according to their initial leakage current measurements. The one showing a \textit{normal spectrum} was measured with an initial \ileak\ of 37 pA and 3.2 nA at $-20^\circ$C and $+25^\circ$C respectively, while the values for the detector with an \textit{abnormal spectrum} were 13 pA and 1.1 nA.}
\label{remoue:fig6}
\end{center}
\end{figure}

For each detector, an $^{241}$Am spectrum was recorded, with an acquisition time of 180 seconds. The detectors are irradiated from the cathode side (Pt contact), for the best charge collection efficiency, as electrons have a better $\mu\tau$ product than holes ($\mu$ and $\tau$ are respectively the mobility and the mean lifetime of the charge carriers inside the semiconductor) \cite{2001ITNS...48..950T,1998NIMPA.406..287W}. We use a common multi-channel analyzer, with the charge signal being integrated in the Amptek A250 pre-amplifier and shaped by an Ortec 572 amplifier, with a shaping time of 2 $\mu$s, before encoding. Through this test, we want to check that the detectors having passed the acceptance test for the ASIC are also good spectrometers. The mean energy resolution at 59.6 keV ($-20^\circ$C, -600 V) is about 1.8 keV, and shows little dispersion around the mean (see Figure \ref{remoue:fig5}). The indicated energy resolution (in terms of Full Width at Half Maximum, FWHM) accounts for the intrinsic resolution of the detector together with the chain noise, which is dominating here ($ENC_{\rm chain}$ = 135 $\pm 10$ electrons, which represents 1.4 $\pm 0.1$ keV in resolution, at a 2 $\mu$s peaking time). Of the good detectors showing low leakage current, about 10\% show abnormal spectra (see Figure \ref{remoue:fig6}). For those, the most likely explanation is a degraded experimental set-up during the measurements. Yet, an abrupt increase of the leakage current during the test cannot be excluded (see section 3.2), since the \ileak\ measurements are performed 30 seconds after the bias voltage is switched on. Further investigations will be conducted on these detectors, but for the time being, they are excluded from the selection of flight detectors. This leaves us with 80\% of the 5000 detectors, passing both tests successfully. 

\subsection{Influence of the Pt layer thickness on the ECLAIRs detection efficiency at 4 keV}
\begin{figure}
\begin{center}
\centerline{\includegraphics[width=.44\textwidth]{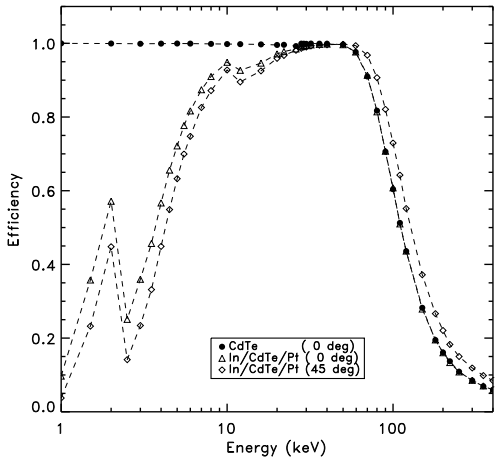}}
  \caption{Simulated absorption efficiency of an In/CdTe/Pt detector, taking into account an irradiated Pt layer (250 nm thickness), compared with a naked CdTe. Considering the detector, we compared a beam perpendicular to the surface (empty triangles) with a beam impacting with a 45 degrees angle (diamonds), which is the maximum incident angle on ECLAIRs detection plane.}
\label{remoue:fig7}
\end{center}
\end{figure}

In parallel of this test campaign, a dedicated destructive physical analysis (DPA) has been carried out on 3 detectors to measure the thickness of the Pt layer, in which X-ray photons up to 30 keV are absorbed (Figure \ref{remoue:fig7}). A 100 nm difference in the Pt thickness would lead to $\sim 10$\% difference in the absorption efficiency at 4 keV. The overall Pt layer was measured to be 200-300 nm thick, which is in the range of the value supplied by Acrorad. This limits the detector absorption efficiency to 45-55\% at 4 keV, as estimated for a 250 nm Pt layer (see Figure \ref{remoue:fig7}). The scientific impact for ECLAIRs has been evaluated through Monte Carlo simulations by Godet et al. \cite{2009NIMPA.603..365G}. The DPA also revealed some minor irregularities present on the edges and corners of the detectors, resulting from the cutting process. 

Precise XRDPIX quantum efficiency measurements (below 20 keV) will be carried out using the SOLEX (\textit{Source Of Low-Energy X-rays}) facility \cite{2004NIMPA.516..594B}.

\section{Time stability of Schottky CdTe detectors}
The time stability of detectors is one of the main issues for practical applications, as for ECLAIRs. Indeed, Schottky-type CdTe detectors suffer from instability induced by a long term application of the bias voltage. 
This results in spectroscopic degradation, the so-called polarization phenomenon \cite{1974ITNS...21..322M,2006JaJAP..45.8842T,2009ApPhL..94j2113C}, as well as an increase in the leakage current with time, due to the Schottky barrier height lowering \cite{2006JaJAP..45.8842T,2009ApPhL..94j2113C}.
Therefore we investigated these instabilities, and their implications for ECLAIRs, first to define the duration of the high voltage switch-off cycle in flight, and secondly to improve the selection criteria in order to preserve a 4 keV low-energy threshold.

\subsection{Stability of spectroscopic performance}
\begin{figure*}[!t]
\begin{center} \centerline{\includegraphics[width=.33\textwidth]{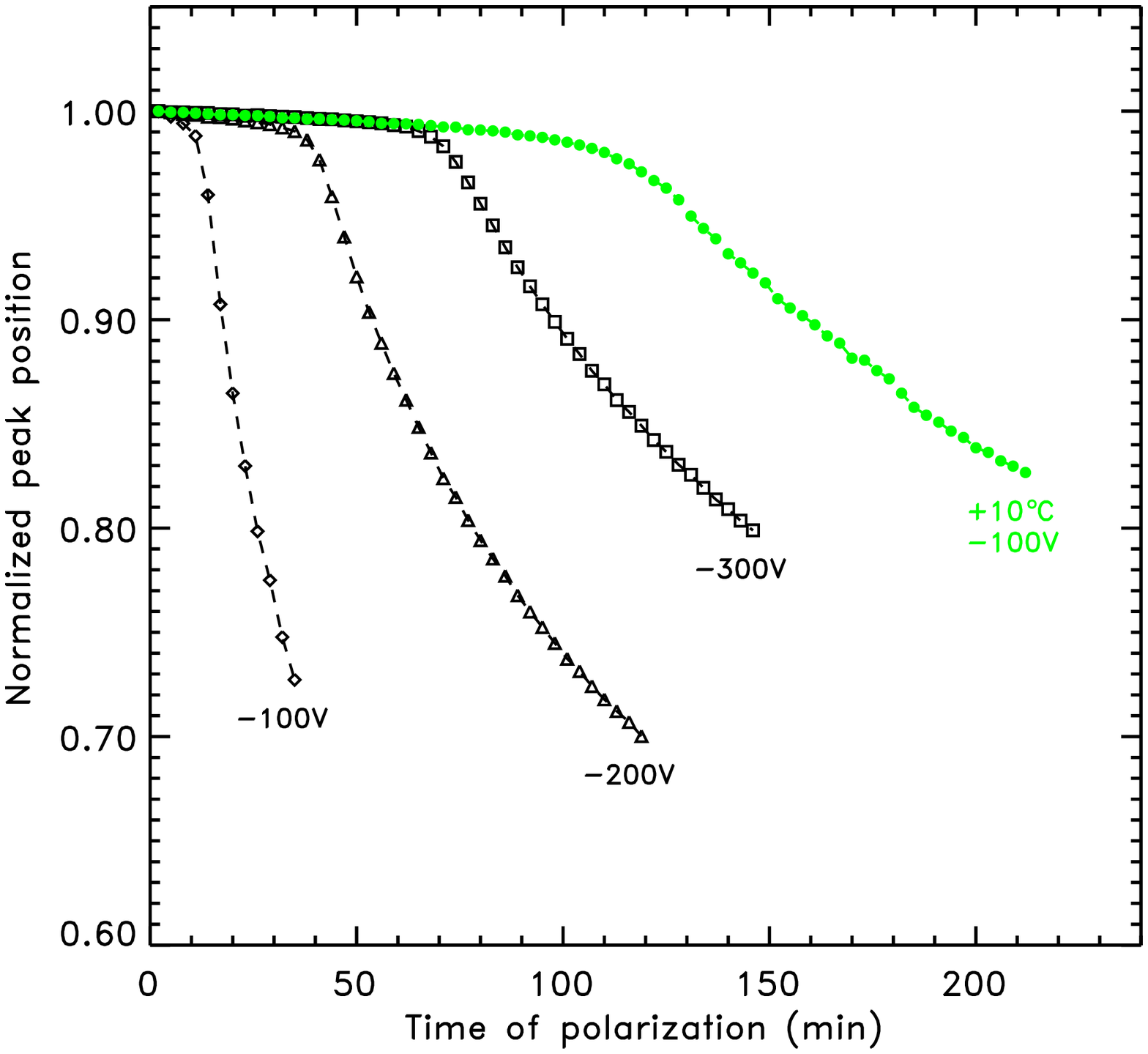}\includegraphics[width=.33\textwidth]{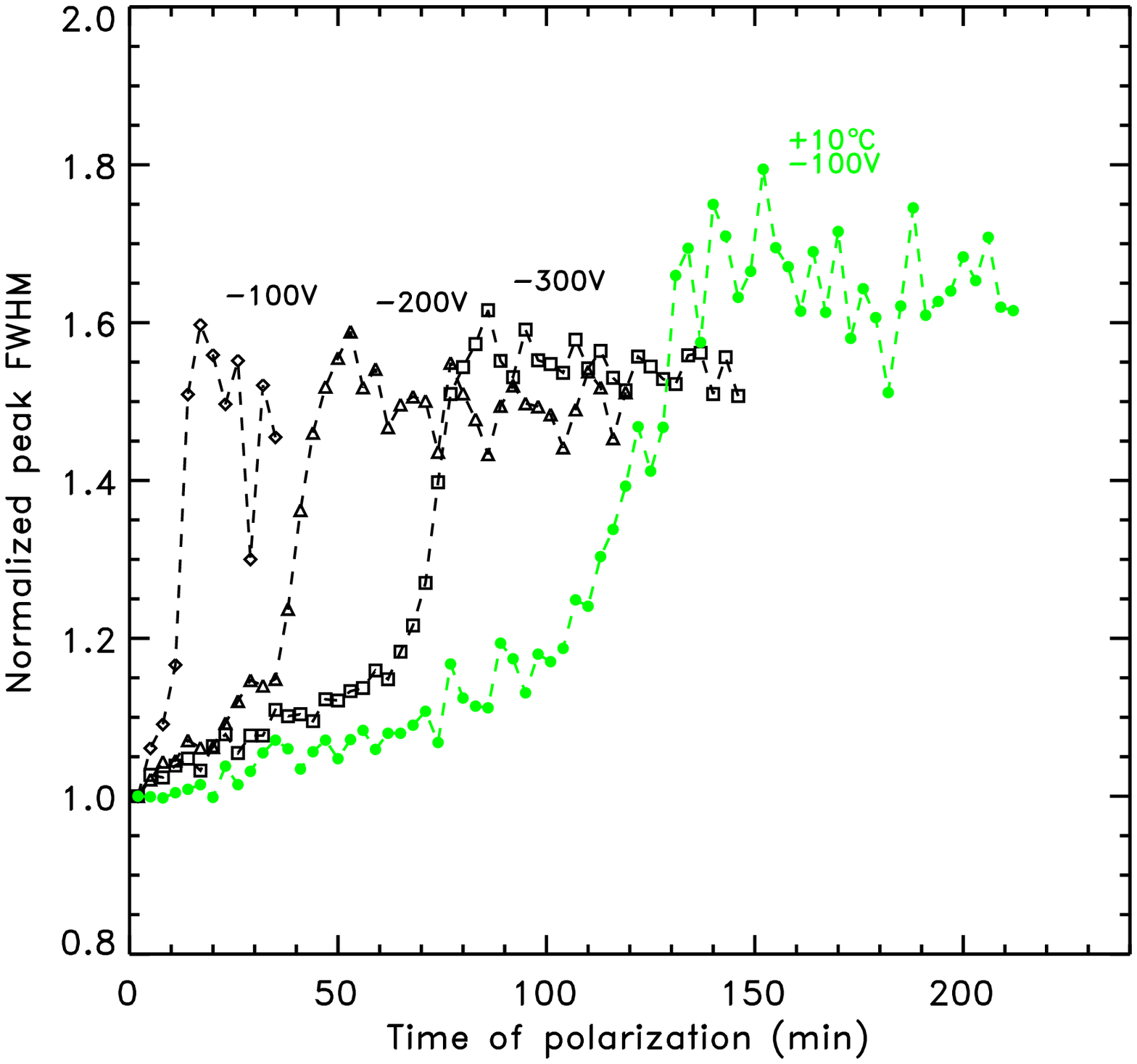}\includegraphics[width=.33\textwidth]{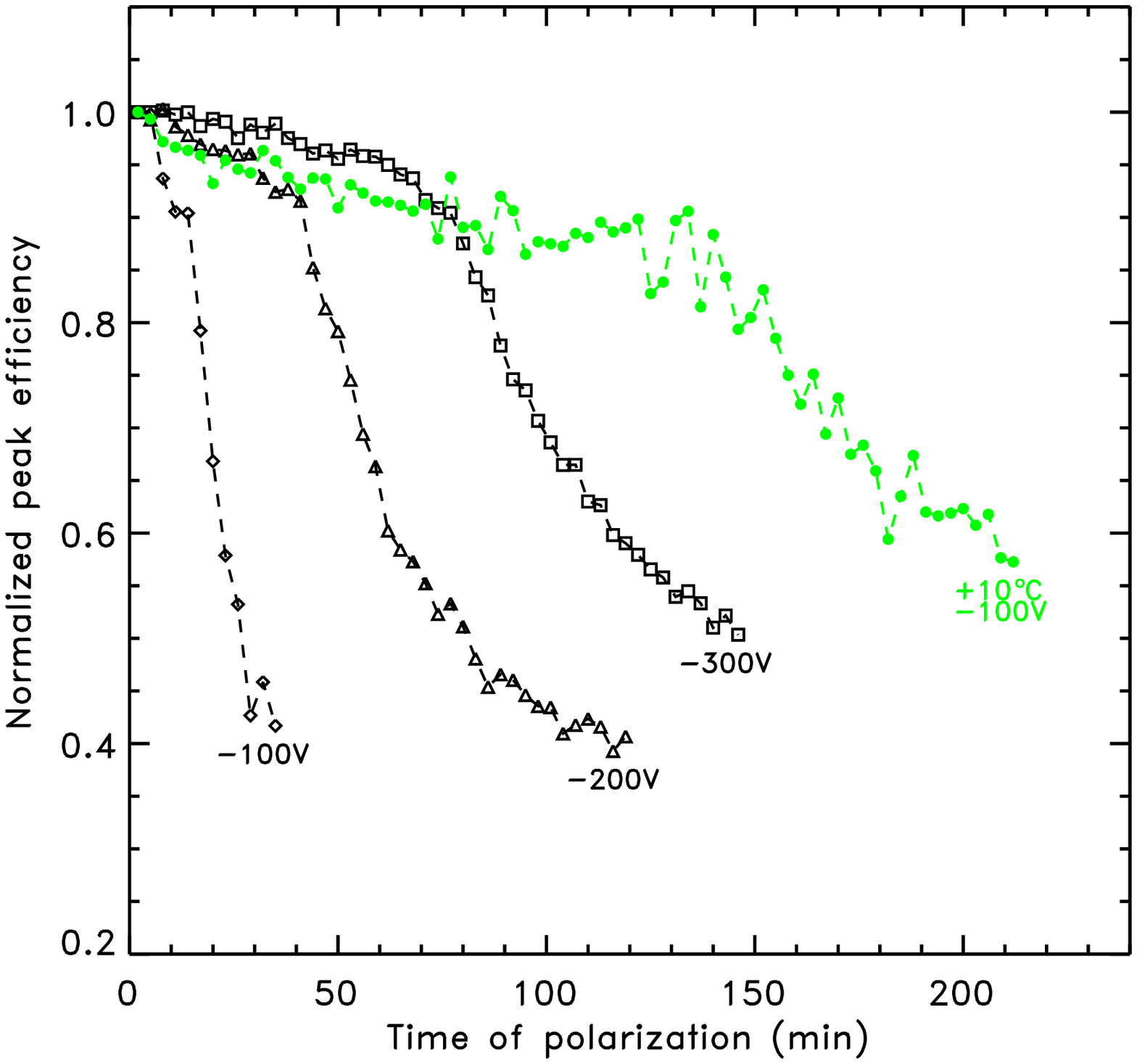}}
  \caption{Impact of the polarization effect on spectroscopy for one representative detector, considering the $^{241}$Am 59.6 keV line, at $+25^\circ$C and various voltages (-100 V, -200 V, -300 V): degradation of the gain with time (left), evolution of the energy resolution (middle) and degradation of the photo-peak efficiency (right). The values measured at $+10^\circ$C under a 100 V reverse bias are also plotted. Data are normalized to the first measurement.}
\label{remoue:fig8}
\end{center}
\end{figure*}

The polarization phenomenon manifests itself mainly by degraded spectroscopic performance with time, i.e. a decrease in quantum efficiency, a degradation of the energy resolution and a shift of the photo-peak towards lower energies. This effect is known to occur more rapidly at high temperatures and low bias voltages \cite{2006JaJAP..45.8842T,1998ITNS...45..428M}. This is illustrated in Figure \ref{remoue:fig8} where three quantities, characterizing the photo-peak at 59.6 keV (its position, width and efficiency), are plotted as a function of time. 
The initial phase is characterized by a smooth variation, until the time $t_p$ when a dramatic degradation occurs. It is worth noting that the polarization phenomenon begins as soon as the bias is turned on. Shortly after $t_p$, the peak FWHM reaches a maximum value; while the photo-peak efficiency shows hints for saturation for $t > 2t_p$. It has been shown that the polarization phenomenon begins accelerating as soon as the electrical field becomes negligible at the cathode \cite{2009ApPhL..94j2113C}, thus implying a decrease of the depletion depth inside the semiconductor. The observed efficiency saturation suggests that the reduction of the depletion depth slows down and stops at a certain thickness (here, below the 1 mm thickness). At -200 V, the quantum efficiency droped by more than 60\% due to the polarization effect.

Studying the 59.6 keV photo-peak position evolution, we determined the time to polarization $t_p$, as well as the velocity ($v_{\rm shift}$) of the smooth initial shift, in various configurations. The results, based on the study of 16 of our flight selected detectors, are listed in Table \ref{remoue:table1}. A linear fit of the initial phase gives $v_{\rm shift}$, while an exponential fitting characterizes the abrupt evolution, with $t_p$ mathematically found at the intersection of both curves. 
Because the dramatic effect appears sooner at -100 V, we have less points recorded before $t_p$ and thus the measurement of $v_{\rm shift}$ is less accurate at this low bias voltage.
The small dispersion suggests that all detectors behave in a similar way. This is reinforced by scaling the line centroid profile to the same $t_p$, between -200 and -300 V at $+25^\circ$C for the 16 detectors, where a very good superimposition is found. A linear increase of the time to polarization with the bias voltage was also observed by Toyama et al. \cite{2006JaJAP..45.8842T}, who studied a 0.5 mm Al/CdTe/Pt Schottky-type detector at $+20^\circ$C in the -50 to -150 V range. We calculated a scaling factor of $\sim 16.3$ s.$\rm V^{-1}$ at $+25^\circ$C for $t_p$. This implies that the spectroscopic performance of our detectors will degrade after about 150 minutes under -600 V at $+25^\circ$C, if the linear trend is still valid at such a high voltage. 

The situation will improve for ECLAIRs, because of the lower temperature and higher voltage applied, as illustrated in Figure \ref{remoue:fig8} at -100 V for a cooler temperature ($+10^\circ$C). Further tests will be performed to complete these preliminary results, and try to define $t_p$ and $v_{\rm shift}$ at $-20^\circ$C and -600 V. All this justifies the needs for ECLAIRs to switch off the high voltage regularly, before $t_p$ is reached or before the slow distorsion of the gain becomes too important that it will be necessary to apply an on-board correction.

\begin{table}[!h]
\begin{center}
\begin{tabularx}{7.5cm}{p{1.0cm} p{1.3cm} p{1.6cm} p{2.5cm}}
\hline
\small{T ($^\circ$C)} & \small{Bias (V)} & \small{$t_p$ (min)} & \small{$v_{\rm shift}$ (eV/min)} \\
\hline \hline
\small{25} & \small{-100} & \small{11 ($\pm$ 1}) & \small{-64 ($\pm$ 5)} \\
\small{25} & \small{-200} & \small{36 ($\pm$ 2)} & \small{-16 ($\pm$ 1)} \\
\small{25} & \small{-300} & \small{66 ($\pm$ 4)} & \small{-7 ($\pm$ 1)} \\
\hline
\small{10} & \small{-100} & \small{107 ($\pm$ 6)} & \small{-7 ($\pm$ 1)} \\
\hline
\end{tabularx}
\caption{Mean time to polarization ($t_p$) and mean velocity of the 59.6 keV photo-peak ($v_{\rm shift}$) shifting towards lower energies before the dramatic decrease for $t > t_p$ in various configurations. The error bars represent the standard deviation over 16 detectors.}
\label{remoue:table1}
\end{center}
\end{table}

Complementary to this study, 54 detectors have been reverse biased at 600 V during 100 hours at $-20^\circ$C while exposed to an $^{241}$Am source. No distortion in the X-ray spectrum occurred between the first and last spectra acquired ($v_{\rm shift}$ $< 10^{-3}$ eV.$\rm h^{-1}$), indicating a good stability of the Schottky CdTe detectors under these conditions. This result is in line with a few day stability measured at $-10^\circ$C and -400 V by Matsumoto et al. on a 0.5 mm thick detector \cite{1998ITNS...45..428M}, and the 4 day stability extrapolated by Cola et al. for a 1 mm thick diode operated at $-10^\circ$C and -500 V \cite{2009ApPhL..94j2113C}. The 100 hour timescale is thus too short to determine $t_p$ at $-20^\circ$C. Moreover, if we assume that all detectors behave in a similar way, no gain correction due to the polarization effect is to be needed within 100 hours, since the energy-to-channel conversion will be 250 eV.$\rm channel^{-1}$ in flight. Therefore, it might be possible to define an in-flight operation duration between two switch-off longer than the currently expected baseline of a day.

The depolarization and the associated timescale, i.e. the time needed to recover full performance by switching off the high voltage, is also an issue for ECLAIRs. The orbital constraints of the SVOM mission give the opportunity to depolarize the detectors during the deepest South Atlantic Anomaly passages, which shall last about 15 minutes. The 16 detectors studied at $+25^\circ$C recovered their initial characteristics in less than 1 minute, but this time might be longer at cooler temperature.
Because of the many implications for ECLAIRs, the time to polarization ($t_p$), the gain evolution ($v_{\rm shift}$), and the time needed for correct depolarization of the detectors, will be further investigated during the XRDPIX long term performance assessments at $-20^\circ$C and -600 V.

\subsection{Time stability of the leakage current}
\begin{figure}[!h]
\begin{center}
\centerline{\includegraphics[width=.45\textwidth]{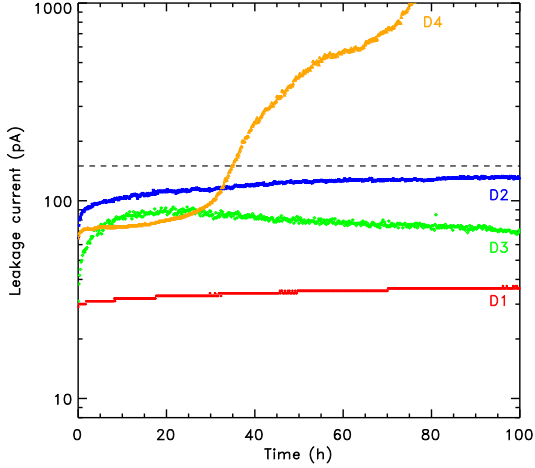}}
  \caption{Time evolution of the leakage current of a few selected detectors, representative of the main classes found. We used ECLAIRs in-flight operating conditions for this test, i.e. $-20^\circ$C and 600 V reverse bias. Most detectors show extremely smooth behaviors, as expected for Schottky-type detectors, over the 100 hours timescale (as D1 and D2). A few detectors show erratic behaviors, with the leakage current increasing significantly with time, during the first hours (as D3) or by major steps (as D4). The 150 pA limit is shown for indication (dashed line).}
\label{remoue:fig9}
\end{center}
\end{figure}

The time stability of the detectors leakage current impacts on the overall performance of each XRDPIX elementary module, which detection threshold should remain below 4 keV on a minimum timescale of a day. Therefore, we performed a long duration test at $-20^\circ$C and -600 V for the 54 detectors. The results are shown in Figure \ref{remoue:fig9}.
80\% of the detectors tested show extremely smooth behaviors over the 100 hour timescale (the curves labelled D1 and D2 in Figure \ref{remoue:fig9} illustrate this profile, with D1 representing the most common case). For such stable detectors, the increase of \ileak\ is limited to less than 40 pA over a timescale of 100 hours. This is equivalent to an increase of $\sim 10$ pA after 24 hours. If we select detectors showing low $I_{\rm leak}$, i.e. typically below 100 pA (as for D1 and D2), this shall ensure stable performance over more than a day, with the requirement of a maximum \ileak\ less than 150 pA fulfilled.

However, some detectors presenting low values behave more erratically with time, as illustrated by D3 or D4 in Figure \ref{remoue:fig9}. 8\% of the detectors tested here show a rapid increase of \ileak\ within a few hours (as for D3), reaching a value a few times higher than the initial one. 
We also found a few detectors (2 over 54 detectors) showing a more unusual behavior (illustrated by D4 in Figure \ref{remoue:fig9}). For those detectors, \ileak\ remains steadily below 150 pA until $\sim 30$ hours, then they show a major increase of \ileak\ (and later on after $\sim 70$ hours). 
Could this happen before switching off the high voltage, this would lead to an increase of the low energy threshold.
In the more extreme case where \ileak\ becomes more than $\sim 2$ nA, it may even saturate the ASIC \cite{2009ITNS...56.2351G}. This would result in the loss of an XRDPIX channel, and therefore a decrease in the effective area.

In fact, the Schottky barrier height is known to lower down with time, thus leading to an increase in the leakage current \cite{2006JaJAP..45.8842T,2009ApPhL..94j2113C}. This is due to an electrical field increase at the anode, induced by an increasing number of ionized acceptors at the Schottky interface \cite{2009ApPhL..94j2113C}. At the same time, the increase of \ileak\ leads to an acceptor ionization decrease inside the bulk, as less carriers are available for ionization \cite{2009ApPhL..94j2113C}. Therefore, there are two competing effects that may explain the slow and smooth increase of \ileak\ on the long run, as illustrated by D1 and D2 in Figure \ref{remoue:fig9}. 
However, we have no satisfying explanation for the decrease in \ileak\ observed for D3, but it may have an extrinsic origin. 
For the most extreme case (D4), the leakage current brutally increases by steps. We might be tempted to interpret this behavior as related to the polarization effect. Yet, during the 100 hour spectroscopic test, no evidence was found for the dramatic degradation.

\begin{figure}[!h]
\begin{center}
\centerline{\includegraphics[width=.45\textwidth]{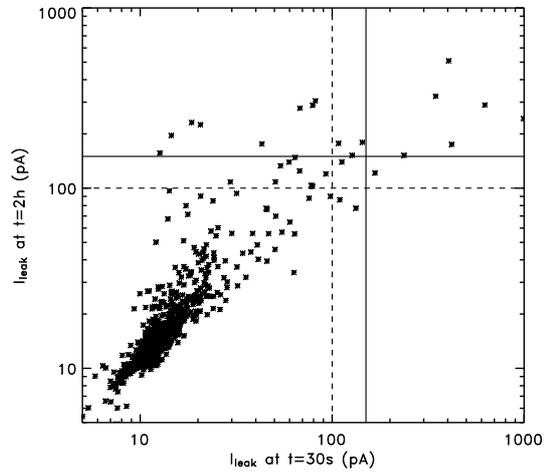}}
  \caption{Leakage currents measured after 2 hours versus the initial \ileak\ (at t=30s), under a -600 V bias voltage and a temperature of $-20^\circ$C, for 1000 detectors. The 150 pA (continuous line) and 100 pA (dashed line) references are plotted for indication. }
\label{remoue:fig10}
\end{center}
\end{figure}

We recently added a 2 hour measurement of the leakage current at $-20^\circ$C and -600 V, in order to estimate the \ileak\ evolution of each individual detector, and eliminate the ones which show a significant increase over this timescale. The preliminary results based on 1000 detectors is shown in Figure \ref{remoue:fig10}. In most cases, the initial increase ends within about 30 minutes, then the leakage current smoothly increases at an average rate of $\sim 2$ pA over 2 hours. This is larger than the value found during the 100 hour test because the timescale is much smaller, and thus the estimation for one day is less accurate. Taking into account the increase in leakage current with time, and allowing for a $\sim 50$ pA margin based on these measurements, we modified our selection criteria to consider only detectors showing a stabilized \ileak\ within 2 hours less than 100 pA. We found 1.9\% of the detectors tested at the moment failing this criterion. 
Those peculiar detectors show an abrupt increase in leakage current during the first 10 to 15 minutes, and then show a smooth behaviour within about 30 minutes as for the other detectors.

\begin{figure}[!h]
\begin{center}
\centerline{\includegraphics[width=.45\textwidth]{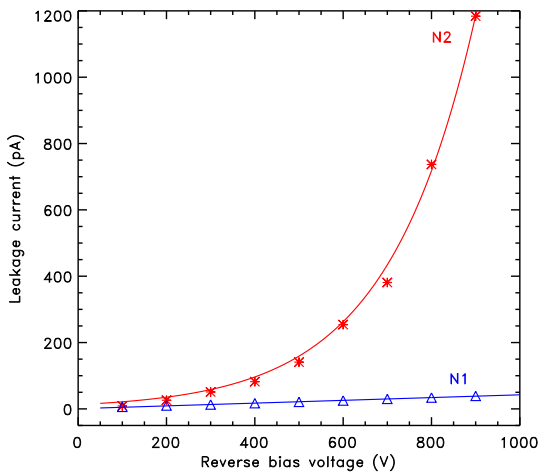}}
  \caption{Leakage current measured at $-20^\circ$C versus reverse bias voltage. The N1 and N2 detectors are representative samples of two detector populations selected from the 2 hour test at -600 V. N1 showed a smooth increase in $I_{\rm leak}$, remaining much lower than 100 pA, while N2 showed a sudden increase in $I_{\rm leak}$, its value exceeding 100 pA after 20 minutes. The two solid lines correspond to the best fits assuming a linear fit for N1 and an exponential function for N2, respectively.}
\label{remoue:fig11}
\end{center}
\end{figure}

Complementary measurements were performed on two sample detectors: one showing a smooth increase in \ileak\ during the 2 hour test at -600 V (labelled N1 in Figure \ref{remoue:fig11}), and one showing a sudden increase in \ileak\ after the first measurement, its value exceeding 100 pA after 20 minutes (labelled N2 in Figure \ref{remoue:fig11}). Their I-V characteristics measured at $-20^\circ$C are shown in Figure \ref{remoue:fig11}. 
The leakage current from N1 shows a linear increase with the reverse bias voltage (from -100 V to -900 V), with an apparent resistivity of 383 $\pm9$ G$\Omega$.m. This is the case of a detector that is biased to a voltage far enough from its barrier breakdown voltage. The leakage current from N2 shows instead an exponential increase with the reverse bias voltage. Such a behavior is reminiscent of that observed when a detector is biased at a voltage close to its breakdown point. This is
consistent with the sudden increase in leakage current observed for N2 during the 2 hour test. If so, the N2 detector then presents a barrier breakdown voltage lower than that for the N1 detector. This may be due to some defects in the Schottky contact (edge or/and surface defects, induced during the cutting process of the detectors for instance). This outlines the importance to perform a 2 hour leakage current test at -600 V, in order to discard detectors showing similar properties to those observed for N2.

If we assume that the first 5000 detectors tested behave in a similar way, which seems to be consistent with all the complementary tests performed and the $I_{\rm leak}$-values measured at t=30s, we may consider that about 2\% of the detectors tested will not remain below 100 pA within 2 hours. This leaves us with 78\% of the detectors already tested selectable to build the XRDPIX flight modules.

\subsection{Application of a charge accumulation model}
Toyama et al. \cite{2006JaJAP..45.8842T} introduced a method which allows, through leakage current measurements, to determine the time when the polarization phenomenon becomes obvious in spectroscopy (i.e. $t_p$). This method is based on a charge accumulation model, which involves deep acceptor levels \cite{1974ITNS...21..322M}. Some parameters of the deep acceptors such as the concentration N$_T$, the hole detrapping time $\tau$, and the activation energy, are extracted. This model also defines the parameter $\alpha$, homogeneous to a distance, which links the barrier height lowering to the electrical field at the anode. 
We used our long duration leakage current and spectroscopic measurements at $+25^\circ$C to investigate the applicability of the model to our data. If it is conclusive, then this would allow us to predict $t_p$ at $-20^\circ$C using fewer and faster measurements than the spectroscopic ones.

In the conventional charge accumulation model \cite{2006JaJAP..45.8842T}, all deep acceptors are assumed to be empty before the bias voltage is applied. In the modified model \cite{2006JaJAP..45.8842T} it is considered that, because of the bending of the bands near the Schottky contact, deep levels in the vicinity of the interface (i.e. at a distance between 0 and $\lambda$ from the anode) are already occupied by electrons before the bias voltage is turned on. In the conventional model, $\lambda$ is equal to zero. This induces a different expression for the electrical field inside the bulk.
Toyama et al. \cite{2006JaJAP..45.8842T} found a good agreement between the reduction of the depletion depth deduced from the modified model and the observed gain loss due to the polarization effect. 

\begin{figure*}[!t]
\begin{center}
\centerline{\includegraphics[width=.48\textwidth]{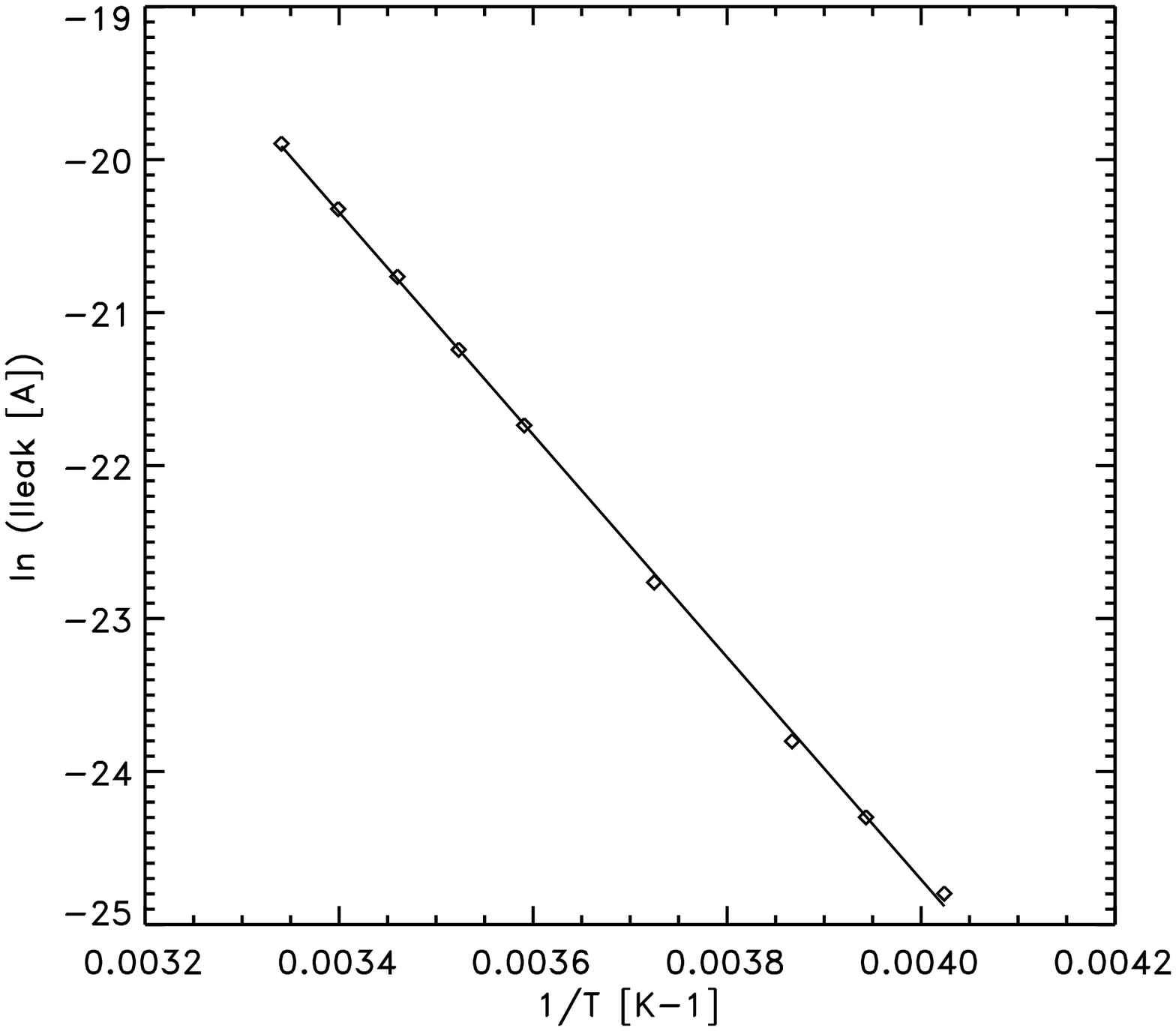}\includegraphics[width=.48\textwidth]{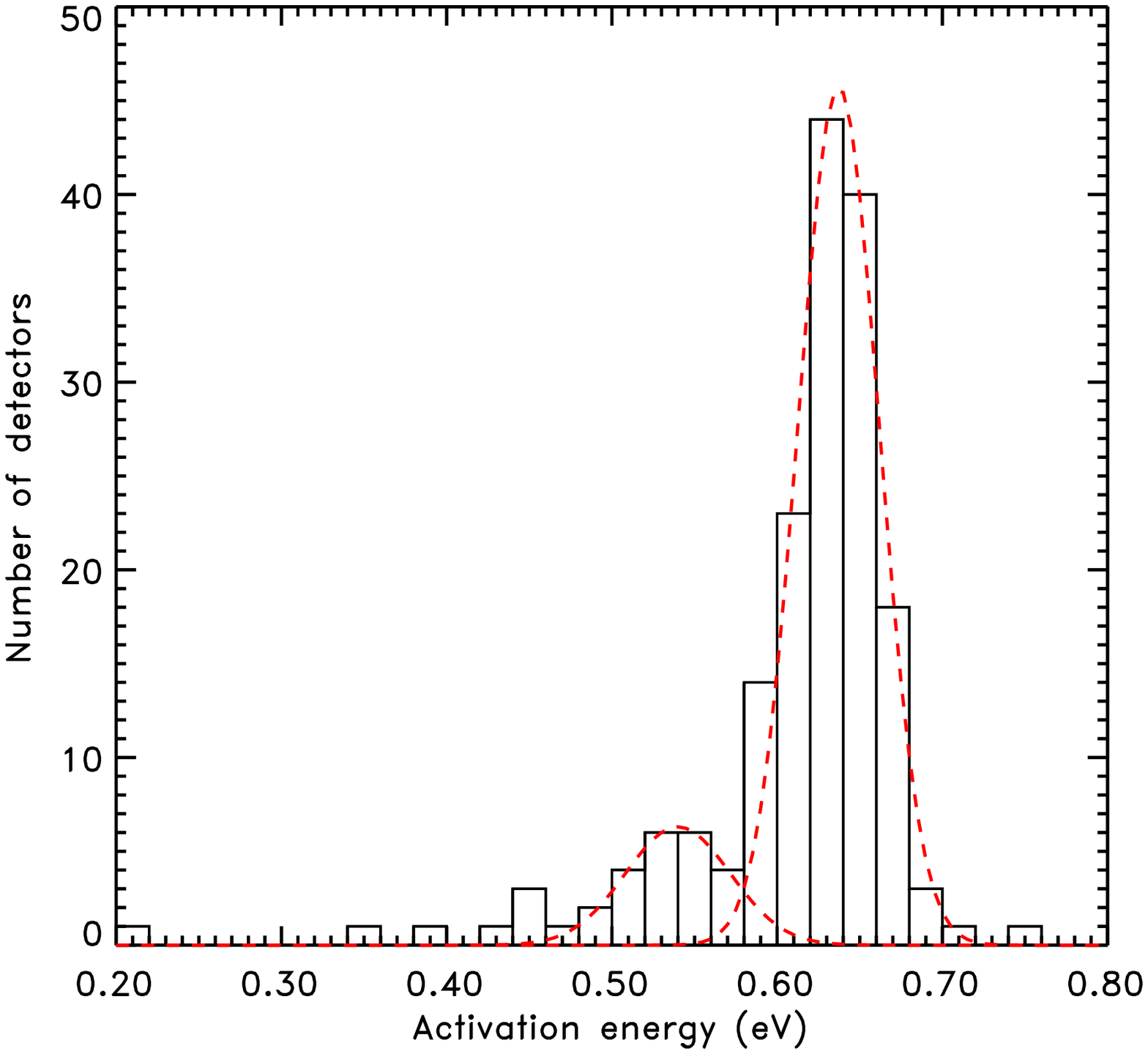}}
  \caption{(a) Linear fitting of the leakage current as a function of 1/T, for one representative detector. The measurements were performed at -600 V, 2 minutes after applying the voltage, and the temperature was stabilized ($\pm 0.2^\circ$C). (b) Distribution of the activation energy for the 170 detectors studied. When fitted with two gaussians, the centroid energy of the main population is at 0.64 eV, while the centroid of the second smaller population is found at 0.54 eV.}
\label{remoue:fig12}
\end{center}
\end{figure*}

Following the same method as described in \cite{2006JaJAP..45.8842T}, we derived the values for $\alpha$, N$_T$, $\lambda$ and $\tau$ for the 16 detectors already studied in section 3.1, using leakage current measurements performed during a 4 hour test at $+25^\circ$C, with a reverse bias ranging from 100 to 300 V. First, we found $\alpha$ = 73 $\pm4$ nm and N$_T$ = 3.0 $\pm 0.6$ $\times$ 10$^{11}$ cm$^{-3}$, which are values similar to both the conventional and modified models. This is the first difference with \cite{2006JaJAP..45.8842T}, which found a higher $N_T$-value with the modified model than with the conventional one, and a higher $\alpha$-value. Then using the modified model, we calculated a value for $\lambda$ no more than a few angstr$\ddot{o}$ms, which is negligible compared to the detector thickness. This is in line with the conventional model, where $\lambda$ is zero. This results in an average value of $\tau$ = 77 $\pm 24$ min at $+25^\circ$C. 
Injecting these parameters into the electrical field expression \cite{2006JaJAP..45.8842T}, we can define the time $t_w$ when the field becomes zero at the cathode. This calculation is restricted to reverse bias values below the factor q$N_T$$D^2$/$2\epsilon$ which is 266 V in our case. We found $t_w$-values higher than the ones actually measured in spectroscopy ($t_p$); e.g. $t_w$ $\sim 100$ min at -200 V, which is about 3 times $t_p$.

Cola and Farella \cite{2009ApPhL..94j2113C,Farella2009} measured the electrical field profile inside the detector as a function of time. The detector studied is the same as the ones we use, the only difference being the area of 10$\times$10 mm$^2$. They measured an $\alpha$-value of 10.5 $\pm2.1$ nm at different temperatures, which is 10 times lower than the value calculated by \cite{2006JaJAP..45.8842T} and lower compared to our value. Moreover, Cola et al. \cite{2009ApPhL..94j2113C} have shown that first, the electrical field profile has no constant component at t=0 (i.e. when the bias is turned on), meaning that deep acceptors from all the bulk are partially ionized under equilibrium; and secondly, the electrical field at the cathode becomes negligible, but not completely zero.
A model taking into account those experimental features could be more accurate to meet our data than the charge accumulation model alone.

\section{Activation energy of Schottky CdTe detectors}
Deep acceptor levels play a key role in the functional properties of the Schottky-type CdTe detectors \cite{1974ITNS...21..322M,2006JaJAP..45.8842T,2009ApPhL..94j2113C}. Therefore, as a by-product of our extensive testing, we measured the mean activation energy $\delta E_{\rm A}$ of our detectors with unprecedented accuracy. As the crystal used is slightly of p-type \cite{1999NIMPA.436..120F}, we assumed that the leakage current depends exponentially on the inverse of the temperature as: \ileak\ = C $\times \exp(- \delta E_{\rm A}/kT)$, where C is a function of the bias applied, the area of the detector, the charge density on the valence band and the hole mobility \cite{kireev75}. 

\begin{figure*}[!t]
\begin{center}
\centerline{\includegraphics[width=.48\textwidth]{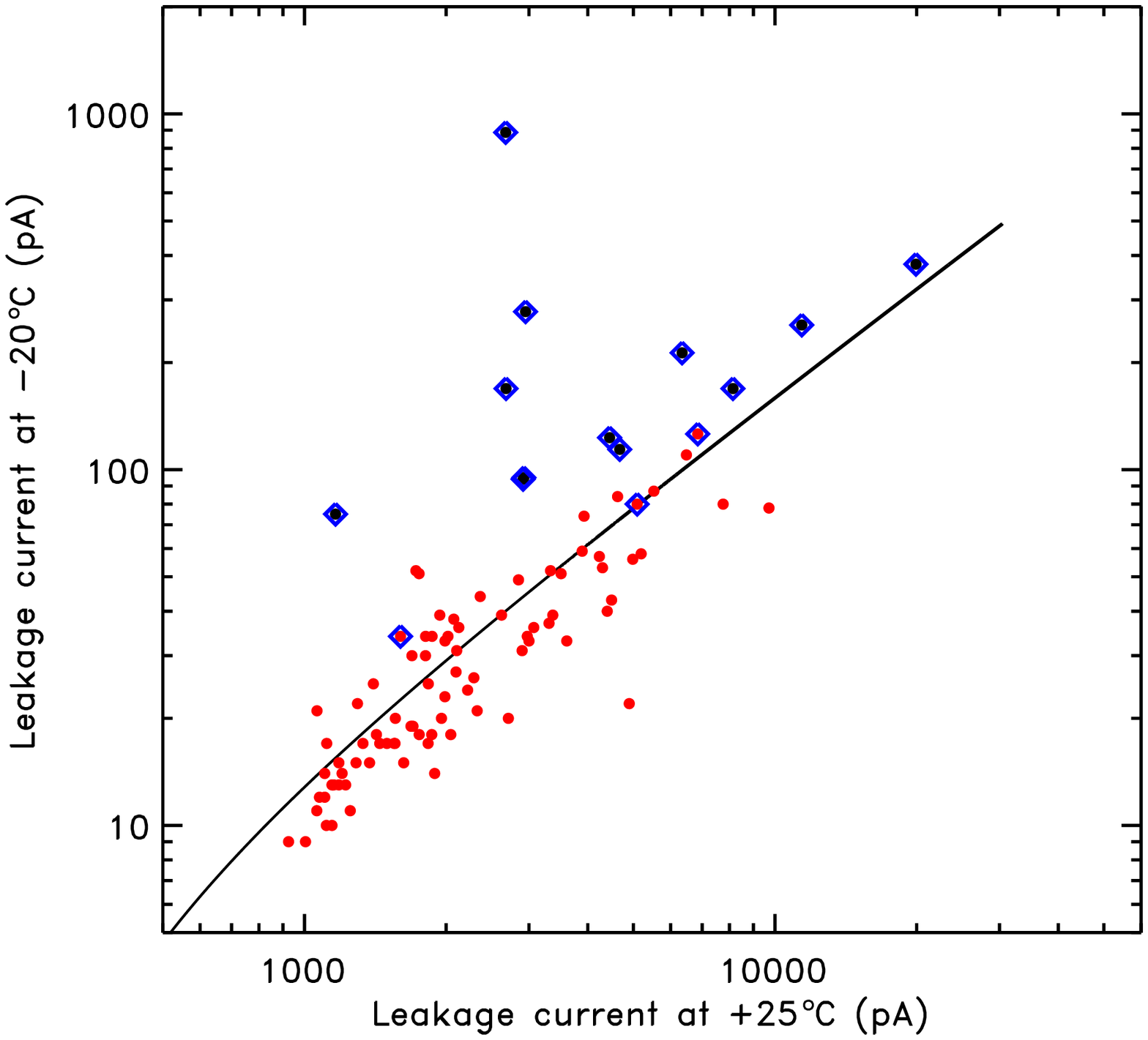}\includegraphics[width=.48\textwidth]{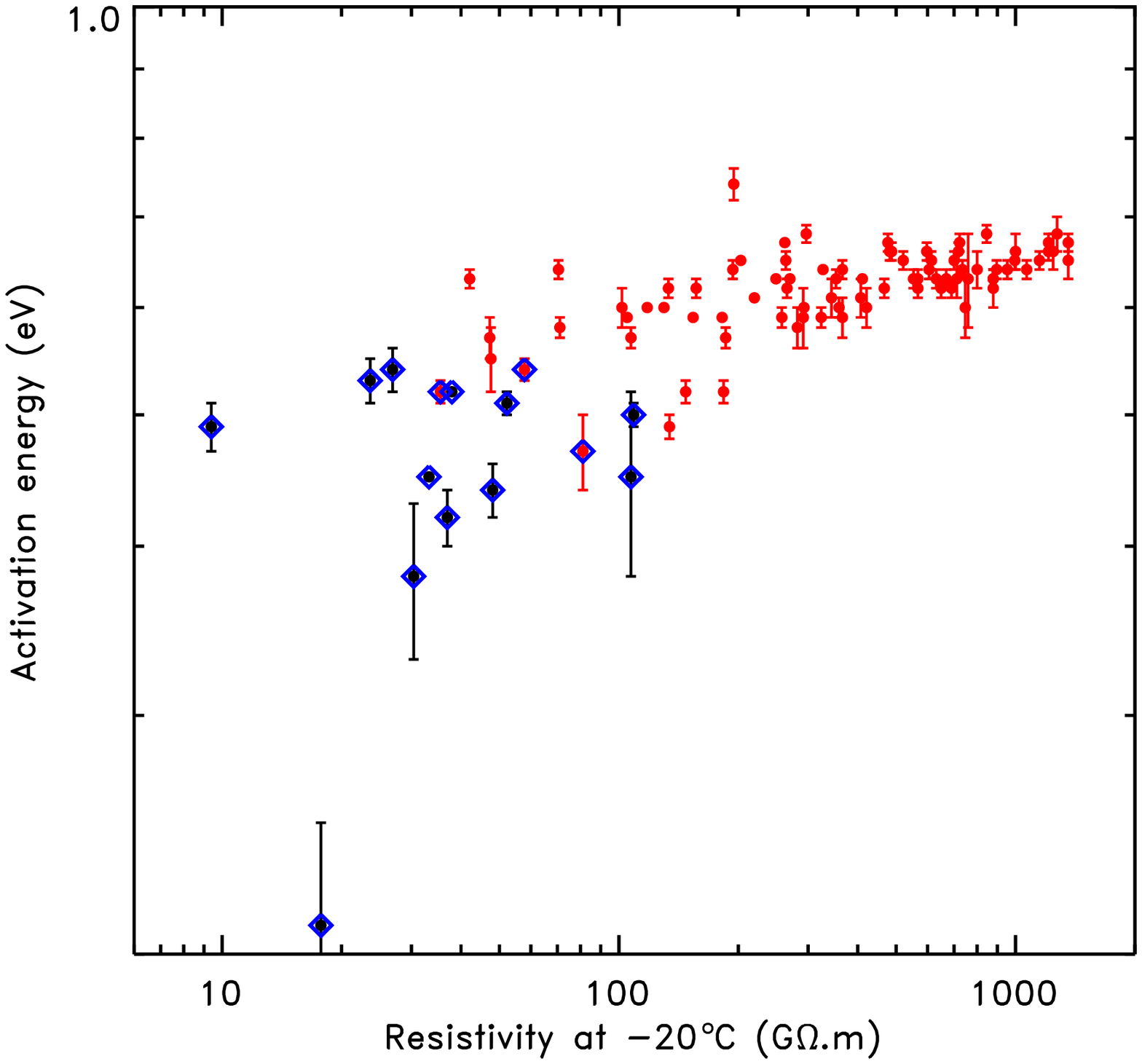}}
  \caption{Study of the leakage current of 100 detectors as a function of temperature and reverse bias voltage. (a) The leakage current measured at $-20^\circ$C against the leakage current at $+25^\circ$C. The linear fit of the main population of detectors is shown for indication (same line as presented in Figure \ref{remoue:fig2}). (b) The activation energy (deduced from leakage current measurements at various temperatures) against the apparent resistivity measured at $-20^\circ$C (deduced from the I-V semi-log characteristics). On both plots, the red points identify the detectors whose \ileak\ remain less than three sigma away from the linear fit, and the diamonds identify detectors which show an activation energy below 0.54 eV together with an apparent resistivity below 110 G$\Omega$.m at $-20^\circ$C.}
\label{remoue:fig13}
\end{center}
\end{figure*}

Using daily measurements, and considering the main population of acceptable detectors (plus signs in Figure \ref{remoue:fig2}), we found values distributed between 0.5 and 0.7 eV, with a mean at 0.64 eV. To improve on the accuracy of the later value, we measured the leakage current increase at temperatures ranging from $-25^\circ$C to $+25^\circ$C for 170 detectors, representative of all the detectors tested i.e. not only the flight selected ones. The linear fit is illustrated on Figure \ref{remoue:fig12} (a). As shown in Figure \ref{remoue:fig12} (b), we obtained a mean activation energy of 0.64 $\pm 0.03$ eV for 85\% of the detectors. A second smaller population is observed at 0.54 $\pm 0.04$ eV. 
The 0.64 eV value is consistent with the ones found by Cola et al. \cite {2009ApPhL..94j2113C} and Toyama et al. \cite{2006JaJAP..45.8842T}, obtained using different approaches. This particular level is often interpreted as being due to cadmium vacancies, and might depend on the thermal treatment during the crystal growth \cite{2003ITNS...50..229A}. The 0.54 eV value, involving fewer detectors, was also previously reported \cite{2003ITNS...50..229A,laasch1995} and interpreted as a donor level present under specific concentration of Cl dopant. This suggests slight modifications during the slow crystal growth process, which might have induced different configurations in the ingot. However, we do not have the information on where each detector come from in the ingot; thus we cannot tell if the detectors showing similar behaviors were adjacent. 

The Cd lattice faults in the crystal, which are present as native defects, have been shown to play an important role in the resistivity and might even be responsible for the polarization phenomenon \cite{2003ITNS...50..229A}. Therefore, we calculated the apparent resistivity of 100 detectors among the 170 detectors, using the linear fit of the I-V semi-log curve as suggested in \cite{1999NIMPA.436..120F}. \ileak\ was measured 2 minutes after applying the bias voltage, ranging from -100 V to -600 V, at $-20^\circ$C. The results are shown in Figure \ref{remoue:fig13}. We found that all detectors away from the main correlation show lower activation energy (typically below 0.54 eV) and lower apparent resistivity (typically below 110 G$\Omega$.m at $-20^\circ$C). When we put those detectors with low activation energy and low apparent resistivity (identified by the diamonds) in Figure \ref{remoue:fig13} (a), we find that 12 detectors out of 15 are located more than three sigma away from the correlation line. Therefore, we have strong indications for intrinsic origins explaining the 10\% detectors which remain away from the main correlation after re-testing in section 2.1.

\section{Conclusions}
After extensive testing of an homogeneous set of more than 5000 Schottky CdTe detectors, foreseen for the detection plane of the X-/Gamma-ray telescope on board the SVOM mission, we can draw the following conclusions, relevant to their in-flight operation and to the requirement for ECLAIRs to detect X-ray photons down to 4 keV. 
\begin{itemize}
\item First, we defined selection criteria for the detectors to be mounted on the XRDPIX flight modules. Under ECLAIRs in-flight operating conditions ($-20^\circ$C and a 600 V reverse bias), they show a low initial leakage current (with a mean value around 20 pA), which remains below 100 pA within 2 hours. They are located less than three sigma away from the main correlation line describing the relation between \ileak\ at $+25^\circ$C and \ileak\ at $-20^\circ$C. These criteria are based on the worse possible hybridization scheme of the detectors with the ASIC IDeF-X, and therefore they are very conservative. Using a standard analog electronics chain, we measured a mean energy resolution of 1.8 keV at 59.6 keV, with the FWHM-value being dominated by the chain ENC ($\sim 135$ electrons rms), i.e. these detectors should intrinsically better resolve the X-ray lines than can be measured here. Yet, we found about 10\% detectors with bad spectra, to be removed from the selection. 
Therefore, taking into account all these criteria, we found that about 78\% of flight pre-selected detectors at the moment. 

\item The absorption due to the Pt layer on top of the 1 mm thick detector determines its efficiency below 20 keV, with about 45-55\% of photons lost at 4 keV. Because the Pt layer was found to be variable, between 200 and 300 nm, precise quantum efficiency measurements will be carried out on a sample of XRDPIX units using the SOLEX facility. A $^{55}$Fe radioactive source, showing a characteristic X-ray line at 5.9 keV, will also be used during the XRDPIX performance assessments.

\item The bias-induced polarization effect has been shown to occur on a very short timescale at high temperatures and low bias voltages, justifying the needs for operating ECLAIRs at $-20^\circ$C and -600 V. We investigated the conventional and modified charge accumulation models described in \cite{2006JaJAP..45.8842T}, using our leakage current measurements at room temperature. Our data were found to be more in agreement with the conventional model. We also derived the time $t_w$ when the depletion depth begins to decrease, and we found that, even if the $t_w$-values give a rough estimation of the $t_p$-values measured in spectroscopy, they are systematically higher.

\item The time stability of the detector performance is essential to preserve the science capability of the instrument ECLAIRs. To do so, we need to define the in-flight operation conditions taking into account the intrinsically unstable nature of the Schottky CdTe detectors. It will be necessary to switch off the high voltage on a regular timescale in order to limit the effect of the polarization phenomenon as well as the constant increase in the leakage current with time. 
The measurements performed on pre-selected detectors, when biased at -600 V and cooled down to  $-20^\circ$C, showed that their spectroscopic performance (quantum efficiency, gain and energy resolution) should be stable on a timescale larger than a day.
The long duration leakage current measurements, performed on a 100 hour timescale, showed that in most cases the $I_{\rm leak}$-values should stay below the maximum 150 pA allowed to maintain a 4 keV low-energy threshold. 
Yet, precise determination of the time to polarization $t_p$ of these detectors at $-20^\circ$C and -600 V is needed in order to determine the nominal duration of operation in orbit, which could be longer than a day following our preliminary tests, as well as the timescale for full recovery of the detector performance when switching off the high voltage. This will be investigated at the XRDPIX level.

\item As a by product of our test campaign, we measured the mean activation energy of 170 Schottky CdTe detectors, representative of all the detectors tested. We found evidence for two distinct populations: the main one centered at 0.64 eV and the second one centered at 0.54 eV. Those two values have been previously reported, albeit over a limited sample of detectors. The 0.64 eV energy activation is often reported \cite{2006JaJAP..45.8842T,2009ApPhL..94j2113C,2003ITNS...50..229A} and is likely to be due to cadmium vacancies, while the less frequent 0.54 eV level might be present under specific concentration of Cl dopant \cite{2003ITNS...50..229A,{laasch1995}}. We showed that the second population centered at 0.54 eV is correlated with: i) an apparent resistivity lower than that computed for the main population (typically below 110 G$\Omega$.m at $-20^\circ$C); ii) a sample of non-selected detectors showing higher $I_{\rm leak}$-values at $-20^\circ$C.

\end{itemize}

\section*{Acknowledgments}
We thank the ECLAIRs CESR team, starting with Roger Pons, the project manager. We are grateful to Carine Amoros, Baptiste Houret, Karine Lacombe, Jacques Land\'e, Wilfried Marty, Gilbert Rouaix and Kevin Sarcelle for developing, maintaining and running the test benches. We thank Olivier Limousin for constructive discussions and suggestions along the preparation of this paper, and Olivier Gevin for his inputs about the ASIC. We thank Wataru Inui from Acrorad Co., Ltd, for his friendly assistance and support during the initial phases of the work presented here. We thank the reviewers for useful and helpful comments. This work is supported by the French Space Agency (CNES) and the French Research Council (CNRS).




\end{document}